\documentclass{amsart}
\usepackage{amsmath,amssymb}

\usepackage{enumerate}

\newtheorem{theorem}{Theorem}[section]
\newtheorem{lemma}[theorem]{Lemma}

\newtheorem{prop}[theorem]{Proposition}

\theoremstyle{definition}

\newtheorem{defn}[theorem]{Definition}

\theoremstyle{remark}

\numberwithin{equation}{section}

\newcommand{\ra}{\rightarrow}

\newcommand{\dfn}{\textbf} 
\newcommand{\mdfn}[1]{\dfn{\mathversion{bold}#1}}

\newcommand{\R}{\mathbb R}
\newcommand{\Rp}{\R^{\geq 0}}
\newcommand{\HH}{\mathbb H}
\newcommand{\Ht}{{\HH^3}}
\newcommand{\iH}{\int_\Ht}
\newcommand{\dx}{\,\textup{d}x}
\newcommand{\dph}{\,\textup{d}\phi}
\newcommand{\dth}{\,\textup{d}\theta}
\newcommand{\dr}{\,\textup{d}r}
\newcommand{\st}                {\,:\,}
\newcommand{\Kij}{K_{p_{ij}}}
\def\k#1#2{k_{#1#2}}
\def\nth#1{\frac{1}{#1}}
\newcommand{\half} {{\nth{2}}}

\newcommand{\Pf}{\mathit{Pf}}

\usepackage[ps,matrix,poly,curve,graph]{xy}
\labelmargin-{1.2pt}  

\newcommand{\TenJ}{%
\begin{xy} 
\xygraph{!{<4pc,0pc>:}
  !P5"A"{~><{@{{}{-}*{\bullet}}} ~>>{_{j_{\xypolynode}}}}
  "A1" -@-_{j_6} "A3"
  "A2" -@-_{j_7} "A4"
  "A3" -@-_{j_8} "A5"
  "A4" -@-_{j_9} "A1"
  "A5" -@-_{j_{10}} "A2"
}
\end{xy}
}

\newcommand{\TenJempty}{%
\begin{xy} 
\xygraph{!{<4pc,0pc>:}
  !P5"A"{~><{@{{}{-}*{\bullet}}}   }
  "A1" -@- "A3"
  "A2" -@- "A4"
  "A3" -@- "A5"
  "A4" -@- "A1"
  "A5" -@- "A2"
}
\end{xy}
}

\begin{document}

\title{Finiteness of Lorentzian 10j symbols and partition functions}

\author{J. Daniel Christensen}
\address{Department of Mathematics \\
University of Western Ontario \\
London, Ontario, Canada }
\email{jdc@uwo.ca}

\begin{abstract}
We give a short and simple proof that the Lorentzian 10j symbol,
which forms a key part of the Barrett-Crane model of Lorentzian 
quantum gravity, is finite.  
The argument is very general, and applies to other integrals.
For example, we show that the Lorentzian and Riemannian \emph{causal}
10j symbols are finite, despite their singularities.
Moreover, we show that integrals that arise in Cherrington's work 
are finite.  Cherrington has shown that this implies that the
Lorentzian partition function for a single triangulation is 
finite, even for degenerate triangulations.
Finally, we also show how to use these methods to prove
finiteness of integrals based on other graphs and other
homogeneous domains.
\end{abstract}



\maketitle

\section{Introduction}
\label{sec:intro}

Spin foams models of quantum gravity express amplitudes by
multiplying together factors coming from the vertices, edges
and faces of a spin foam~\cite{Baez,Perez,Rovelli}.  A spin foam is a 
2-dimensional cell complex whose faces are labelled by 
group representations and whose edges are labelled by 
intertwiners.
The factors that are generally the most physically
interesting, as well as being the ones that are most
difficult to define and compute, are the vertex amplitudes.
These are generally expressed as high-dimensional 
oscillatory integrals, often over non-compact spaces.
These integrals are called 10j symbols.

For example, the vertex amplitude for the Barrett-Crane model of 
4-dimensional Lorentzian quantum gravity~\cite{BC2} is the
Lorentzian 10j symbol.
This is a function that assigns to
ten non-negative real numbers $p_{ij}$, $0 \leq i < j \leq 4$
(thought of as representations of the Lorentz group from the 
principal series) a real number.  
Up to a normalization constant, 
it is defined to be the value of the integral
\begin{equation}
\label{eq:L10j}
\iH \iH \iH \iH \prod_{0 \leq i < j \leq 4} \Kij(d_\Ht(x_i,x_j)) \dx_1 \cdots \dx_4.
\end{equation}
In this expression, $\Ht$ denotes 3-dimensional hyperbolic space,
which can be thought of as $\Ht = \{ (t,x,y,z) \in \R^4 \st
t^2 - x^2 - y^2 - z^2 = 1 \text{ and } t > 0 \}$ with the 
induced Riemannian metric, and
$d_\Ht$ denotes the hyperbolic distance.
For $p > 0$, the kernel $K_p$ is defined by
\[
K_p(d) = \frac{\sin(p\, d)}{p \sinh(d)}.
\]
When $p = 0$, we set $K_0(d) = d/\sinh(d)$.
When $d = 0$, we set $K_p = 1$. 
In~(\ref{eq:L10j}), the point $x_0 \in \Ht$ is fixed, and the value 
of the integral doesn't depend on the choice.
The measure $\dx_i$ denotes the usual hyperbolic measure
(see Lemma~\ref{le:edge} below), i.e., the Riemannian volume form.

In this paper, we give a simple proof that the integral~(\ref{eq:L10j}) is finite.
More precisely, we show that it is absolutely convergent, 
a result originally obtained by Baez and Barrett~\cite{BaBa01}.
The present proof is shorter, avoiding the use of hyperbolic
geometry and various estimates.
Also, the argument extends to kernels that can't be handled
by direct application of the methods of~\cite{BaBa01}.
For example, the Lorentzian and Riemannian causal kernels
have divergences at $d = 0$, and yet the present proof
handles these kernels without any changes.
Moreover, it also applies to kernels of
the form $r^k/\sinh(d_\Ht(x,y))$, $k \geq 0$.
For $k < 1$, these kernels are singular and again
cannot be handled by Baez and Barrett's method.
These kernels appeared in work of Cherrington~\cite{Che05},
where he shows that finiteness of the corresponding 10j symbol
implies that the Lorentzian partition function for a single 
triangulation is finite, even for degenerate triangulations.
This was already known for non-degenerate triangulations~\cite{CPR}.

The 10j symbol is an evaluation based on the graph
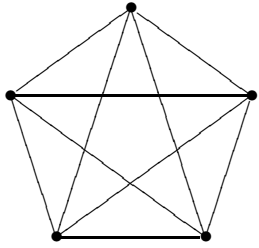
\[ \TenJempty  \]
with one variable per vertex, one kernel per edge, and one integration
omitted. 
The methods of this paper can also be used for evaluations based
on different graphs, by looking at the spanning subtrees of the
graph.
We also treat the case where hyperbolic space $\Ht$ 
is replaced by other homogeneous domains.

\subsection*{Outline}
In Section~\ref{sec:convexity} we give several inequalities 
based on convexity that we will use in the rest of the paper.
In Section~\ref{sec:10j} we present the proof that the Lorentzian
10j symbol is absolutely convergent.
In Section~\ref{sec:other-kernels} we show that this generalizes
to other kernels.
We explore the case of different graphs and different homogeneous
domains in Section~\ref{sec:graphs} and
we give our conclusions in Section~\ref{sec:conc}.

\section{Convexity}
\label{sec:convexity}

We will use the notion of convexity to prove that various
integrals are finite.

\begin{defn}
A function $f \colon \R^n \ra \R$ is \dfn{convex} if for
all $s, t \in \R^n$ and all $\alpha, \beta \in \R$
with $0 \leq \alpha, \beta \leq 1$ and $\alpha+\beta=1$,
we have
\[
  f(\alpha s + \beta t) \leq \alpha f(s) + \beta f(t) .
\]
\end{defn}

It follows that if $t^1, \ldots, t^k$ are points in $\R^n$
and $\alpha_1, \ldots, \alpha_k$ are non-negative real
numbers summing to 1, then
\[
  f(\alpha_1 t^1 + \cdots + \alpha_k t^k) \leq 
    \alpha_1 f(t^1) + \cdots + \alpha_k f(t^k) .
\]

Our main example is the following function.  
Let $a_1, \cdots, a_n$ be fixed positive real numbers,
and define $P \colon \R^n \ra \R$ by the formula 
\[
  P(t) = {a_1}^{\!t_1} \cdots {a_n}^{\!t_n} .
\]
Note that
\[
{a_1}^{\!t_1} \cdots {a_n}^{\!t_n} = \exp(t_1 \ln a_1 + \cdots + t_n \ln a_n).
\]
Since $\exp$ is convex and increasing, and linear functions are convex, 
it follows that $P$ is convex.
(Note that the product of convex functions is not in general
convex.  For example, $t_1 t_2$ is not convex.)

Certain key inequalities follow from the convexity of $P$.
For example, take $n=2$, so $P(t_1,t_2) = a_1^{\!t_1} a_2^{\!t_2}$.
Then $(1,1) = \half(2,0) + \half(0,2)$, so it follows
that $P(1,1) \leq \half P(2,0) + \half P(0,2)$, i.e.,
\begin{equation}
\label{eq:i2}
  a_1 a_2 \leq \half a_1^2 + \half a_2^2 ,
\end{equation}
a familiar inequality.

Or, taking $n=3$, we can deduce that
\[ 
  a_1 a_2 a_3 \leq 
    \nth 3 \left( (a_2 a_3)^{3/2} + (a_1 a_3)^{3/2} + (a_1 a_2)^{3/2} \right), 
\]
since $(1,1,1) = \nth 3 \left( (0,3/2,3/2) + (3/2,0,3/2) + (3/2,3/2,0) \right)$.

For $n=5$, we get the inequality
\begin{equation}
\label{eq:i5} 
  a_1 a_2 a_3 a_4 a_5 \leq \nth 5 \left(
    (a_2 a_3 a_4 a_5)^{5/4} + \cdots + (a_1 a_2 a_3 a_4)^{5/4} \right) ,
\end{equation}
since $(1,1,1,1,1)=\nth5\left((0,5/4,5/4,5/4,5/4)+\cdots+(5/4,5/4,5/4,5/4,0)\right)$.

These inequalities will be very useful below.
We derived them assuming that the $a_i$'s are positive,
but they clearly hold as long as the $a_i$'s are non-negative.

In the above, if there are $n$
factors to begin with, when one is removed, the remaining
factors are raised to the power $n/(n-1)$.  This keeps
the total exponent the same, and uses the fact that
$(1,1,\ldots,1)$ is the barycenter of 
$(0,n/(n-1),\ldots,n/(n-1)),\, \ldots,\, (n/(n-1), \ldots, n/(n-1), 0)$.
The method has more flexibility than this, however.
For example, since 
\[
  (1,1,1,1) = 
    \frac{1}{4} (4,2,1,0) + \frac{1}{2} (0,1,0,1) + \frac{1}{4} (0,0,3,2)
\]
one obtains the inequality
\[
  a b c d \leq 
    \frac{1}{4} a^4 b^2 c + \frac{1}{2} b d + \frac{1}{4} c^3 d^2 
\]
in which the total degrees vary, there are three terms, not four,
and the coefficients are not equal.
We will make use of such ``non-standard'' inequalities when we consider
general graphs in Section~\ref{sec:graphs}.

\section{Lorentzian 10j symbol}
\label{sec:10j}

Now we prove the finiteness of the integral in equation~(\ref{eq:L10j}).
In fact, what we show is that it is absolutely convergent.
The only part of the argument that is specific to the Lorentzian
kernel and hyperbolic space is the following lemma, which uses
the notation from the introduction.

\begin{lemma}
\label{le:edge}
Let $y$ be a fixed point in $\Ht$.  Then the integral
\[
  \iH (K_p(d_\Ht(x,y)))^m \dx
\]
is absolutely convergent for each real number $m > 2$.
Moreover, the value is independent of $y$.
\end{lemma}

\begin{proof}
If
\[
  \iH \frac{|\sin(p\, d_\Ht(x,y))|^m}{p^m \sinh(d_\Ht(x,y))^m} \dx
\]
is convergent,
it is clearly independent of $y$, since it only
depends on the hyperbolic distance from $x$ to $y$,
and for any other point $y'$, there is a distance and volume preserving
diffeomorphism of $\Ht$ sending $y$ to $y'$.

To prove convergence, we work in hyperbolic spherical coordinates,
$(r,\theta,\phi)$, and set $y = 0$.
In these coordinates, 
$\dx = \sinh^2 r \sin \phi \dph \dth \dr$
and $d_\Ht(x,0) = r$.
The integral becomes
\begin{equation*}
  \frac{1}{p^m} \int_0^\infty \int_0^{2\pi} \int_0^\pi 
    \frac{|\sin(p r)|^m}{\sinh^m r} \sinh^2 r \sin \phi \dph \dth \dr
  = \frac{4\pi}{p^m} \int_0^\infty \frac{|\sin(p r)|^m}{\sinh^{m-2} r} \dr .
\end{equation*}
The last integrand is bounded at $r=0$ and decays exponentially
as $r \ra \infty$ (since $m > 2$), and so is convergent.
\end{proof}

\begin{theorem}
\label{th:Lor}
The integral in equation~(\ref{eq:L10j}) is absolutely convergent.
\end{theorem}

\begin{proof}
We will use a concise notation, in which $\k ij$ is short
for $|\Kij(d_\Ht(x_i,x_j))|$.  To make symmetries clear, we write the
subscripts in either order.  We take $x_0 = 0$.

The absolute value of the integrand is
\[
  (\k 01 \k 14 \k 43 \k 32 \k 20)(\k 04 \k 42 \k 21 \k 13 \k 30)
\]
By equation~(\ref{eq:i2}), 
we get the following upper bound on the integrand:
\[
  (\k 01 \k 14 \k 43 \k 32 \k 20)(\k 04 \k 42 \k 21 \k 13 \k 30)
  \leq
  \half \left( (\k 01 \k 14 \k 43 \k 32 \k 20)^2 + 
               (\k 04 \k 42 \k 21 \k 13 \k 30)^2 \right) .
\]
(All of the inequalities discussed here hold pointwise, 
for each $x_1, \ldots, x_4$ in $\Ht$.)
The two terms on the right-hand side are symmetrical, 
so it suffices to show that the first is integrable.  
By equation~(\ref{eq:i5}), we have the inequality
\begin{equation*}
\begin{split}
       (\k 01 \k 14 \k 43 \k 32 \k 20)^2 
 &=    (\k 01^2 \k 14^2 \k 43^2 \k 32^2 \k 20^2) \\
 &\leq \frac{1}{5} \left( 
        (\k 14^2 \k 43^2 \k 32^2 \k 20^2)^{5/4}
      + \cdots
      + (\k 01^2 \k 14^2 \k 43^2 \k 32^2)^{5/4}
      \right) \\
 &=    \frac{1}{5} \left( 
        (\k 14 \k 43 \k 32 \k 20)^{5/2}
      + \cdots
      + (\k 01 \k 14 \k 43 \k 32)^{5/2}
      \right).
\end{split}
\end{equation*}
There are five terms in the last expression.  The first and
the fifth are the same, after reversing order and permuting
the variables.  The same is true of the second and fourth.
So there are just three expressions we must show are integrable:
the first $(\k 14 \k 43 \k 32 \k 20)^{5/2}$,
the second $(\k 01 \k 43 \k 32 \k 20)^{5/2}$ and
the third $(\k 01 \k 14 \k 32 \k 20)^{5/2}$.

For the first case, we order the integrations as follows:
\begin{equation}
\label{eq:case1}
  \iH \iH \iH \iH (\k 14 \k 43 \k 32 \k 20)^{5/2} \dx_1 \dx_4 \dx_3 \dx_2 .
\end{equation}
The innermost integral is
\[
  \iH \k 14^{5/2} \dx_1 ,
\]
which is finite and independent of $x_4$ by Lemma~\ref{le:edge}.
Similarly, the next integral is
\[
  \iH \k 43^{5/2} \dx_4 ,
\]
which produces a constant.  And so on.

For the second case, we use the same order of integration:
\[
  \iH \iH \iH \iH (\k 01 \k 43 \k 32 \k 20)^{5/2} \dx_1 \dx_4 \dx_3 \dx_2 .
\]
This factors into
\[
  \iH \k 01^{5/2} \dx_1 \, 
    \iH \iH \iH (\k 43 \k 32 \k 20)^{5/2} \dx_4 \dx_3 \dx_2 .
\]
The first factor is finite, again by Lemma~\ref{le:edge}, and
the second factor is handled just like~(\ref{eq:case1}).

The third case is similar.
\end{proof}

Note that $|K_p| \leq |K_0|$, and so
\[
  \iH |K_p(d_\Ht(x,y))|^m \dx \leq 
  \iH |K_0(d_\Ht(x,y))|^m \dx =: C_m ,
\]
where $C_m$ depends on $m$ but not on $p$.  
This allows one to obtain explicit upper bounds on the
values of the $10j$ symbol.

\section{Other kernels}
\label{sec:other-kernels}

\subsection{Cherrington's kernel}
\label{ss:cherringtons-kernel}

Fix $k \geq 0$.
In this section, we show that the 10j symbol is finite when
the Lorentzian kernel is replaced by the kernel
\[
Ch^k(d) = \frac{d^k}{\sinh(d)}.
\]
For $k < 1$, this kernel is divergent at $d = 0$, and the
product of ten of these kernels has a complicated singularity structure.
Nevertheless, the analog of Lemma~\ref{le:edge} holds for this kernel,
as long as $2 < m < 3$, and so the proof of Theorem~\ref{th:Lor} goes
through unchanged to show that the integral is convergent.

In more detail, in order for the integral
\[
  \iH (Ch^k(d_\Ht(x,y)))^m \dx
\]
to be absolutely convergent at infinity, we require $m > 2$,
as in Lemma~\ref{le:edge}.
For $k \geq 1$, there are no problems near $d = 0$.
For $0 \leq k < 1$, the integral behaves like $d^{m(k-1)}$
near $d = 0$, so we need to ensure that $m(k-1) > -3$,
that is, that $m < \frac{3}{1-k}$.  This is certainly
the case for $m < 3$, and all that the proof of 
Theorem~\ref{th:Lor} requires is $m = 5/2$.

It then follows from work of Wade Cherrington~\cite{Che05} that
the partition function for a fixed triangulation in the Lorentzian
Barrett-Crane model is finite, 
even in the case of a degenerate triangulation.
This was already known for non-degenerate triangulations~\cite{CPR}.

\subsection{Causal kernels}
\label{ss:causal}

A causal version of the Lorentzian kernel
was defined by Livine and Oriti~\cite{LiOr}:
\[
  LO_p(d) = \frac{e^{\pm i p d}}{p \sinh d} .
\]
Like Cherrington's kernel, this kernel diverges at $d=0$.
At first it was thought that the associated $10j$ symbol
diverges, and motivated by this an alternate causal kernel
was introduced by Pfeiffer~\cite{Pf}:
\[
  \Pf_p(d) = \frac{\sin(p d/2) \, e^{\pm i p d/2}}{p \sinh d} .
\]
In fact, $|\Pf_p(d)| \leq |LO_p(d)| \leq Ch^0(d)/p$,
and so it follows from the previous section that
both of these kernels have convergent $10j$ symbols.

Livine and Oriti, and Pfeiffer define causal versions
of the Riemannian kernel.  Both kernels have divergences,
but the above methods show that they have absolutely
convergent $10j$ symbols.

\section{Generalizations to other graphs, other domains}
\label{sec:graphs}

The method we used for the 10j symbol can be generalized to 
different graphs and different domains.

By a \dfn{homogeneous domain} we mean a Riemannian manifold
such that for points $x, y, x', y'$ in $M$ with 
$d_M(x, y) = d_M(x', y')$, there exists a distance and
volume preserving diffeomorphism from $M$ to $M$ which
sends $x$ to $x'$ and $y$ to $y'$.
Let $K_p \colon \Rp \ra \R$, for $p$ in some indexing set $I$, 
be a family of functions.
Let $G = (V, E, w, p)$ be a connected weighted labelled graph.
$G$ has vertex set $V$, edge set $E$, weight function $w \colon E \ra \R$
and labelling $p \colon E \ra I$ sending $e$ to $p_e$.
Then the \mdfn{$K$-evaluation of $G$ over $M$} is the value of
\[
  \int_M \cdots \int_M \, \prod_{e \in E} K_{p_e}(d_e)^{w_e} 
    \prod_{v \in V'} \dx_v ,
\]
where $V'$ is $V$ with some vertex $v_0$ omitted, $x_v \in M$
for $v \in V$, $x_{v_0}$ is held fixed, and $d_e$ is the 
distance in $M$ between $x_v$ and $x_{v'}$, for $v$ and $v'$
the endpoints of $e$.
This is only well-defined when it is absolutely convergent,
since in the conditionally convergent case the answer could
depend on the order of integration.

If this integral converges, the value is independent of the choice
of $v_0$ and of the choice of $x_{v_0}$, because of the homogeneity of
$M$.

For example, when $G$ is
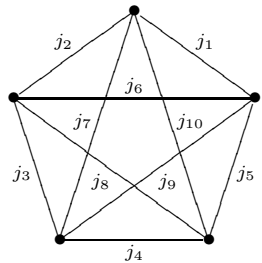
\[ \TenJ \]
with all weights equal to 1, the evaluation is the $10j$ symbol
for the kernels $K_p$.

Note that if a graph has an edge of weight 0, that edge can
be deleted without changing the evaluation.
And if it has parallel edges $e_1$ and $e_2$ between vertices 
$v$ and $v'$, and if $p_{e_1}= p_{e_2}$, then these can be
replaced by a new edge $e$ between $v$ and $v'$ with 
label $p_e = p_{e_1}$ and weight $w_e = w_{e_1}+w_{e_2}$.
We study weighted, labelled graphs up to these equivalences.

If two weighted, labelled graphs have the same vertex set,
they can be added by taking the disjoint union of the edges.
In practice, edges between the same vertices usually have
the same labels, and so one can just add the weights on
the corresponding edges.  
A weighted graph can also be multiplied by a real number $\alpha$ by
multiplying all of the weights by $\alpha$.

For example,
\[
\vcenter{
\xymatrix@=2.5pc{ *{\bullet} \ar@{-}[d]^5_j \ar@{-}[r]_4^k 
          & *{\bullet} \ar@{-}[dl]_3^{\ell} \\ 
            *{\bullet} }
}
+ \, 3 \quad
\vcenter{
\xymatrix@=2.5pc{ *{\bullet} \ar@{-}[d]^1_j \ar@{-}[r]_2^k 
          & *{\bullet} \\
            *{\bullet} }
}
\,=\,
\vcenter{
\xymatrix@=2.5pc{ *{\bullet} \ar@{-}[d]^8_j \ar@{-}[r]_{10}^k 
          & *{\bullet} \ar@{-}[dl]_3^{\ell} \\
            *{\bullet} }
}
\] 
Here we write the labels $j$, $k$ and $l$ as well as the
weights beside the edges.  Below we will also use the
convention that an edge with no explicit weight has
weight 1, and that the labels can be omitted if their
position makes it clear which is which.

\begin{prop}
Fix $M$ and $K_p$ as above.
Let $G, G_1, G_2, \ldots, G_k$ be weighted, labelled graphs 
with the same vertex set.
If $G_1, \ldots, G_k$ are absolutely convergent
and $G$ is in the convex hull of the $G_i$'s,
then $G$ is absolutely convergent.
\end{prop}

\begin{proof}
The edge weights are the exponents in the integrand, so
this follows from the convexity results in Section~\ref{sec:convexity}.
\end{proof}

\begin{lemma}
If $G$ is a tree (a connected, acyclic graph), then the evaluation
of $G$ is absolutely convergent if and only if
\begin{equation}\label{eq:tree}
  \prod_{e \in E} \, \int_M \, K_{p_e}(d_M(x,x_0))^{w_e} \dx
\end{equation}
is absolutely convergent, for any fixed $x_0 \in M$.  
Moreover, when these are absolutely convergent, they are equal.
\end{lemma}

\begin{proof}
Order the integrations from the leaves to the root $v_0$, as was done
in the proof of Theorem~\ref{th:Lor}.
\end{proof}

This suggests a systematic strategy for dealing with a general graph:
express the given graph as a convex linear combination of its
spanning subtrees with weights chosen to be in the range that
makes the integrals~(\ref{eq:tree}) converge.

For example, the theta graph can be expressed in the following
way:
\[
\xymatrix{ *{\bullet} 
\ar@{-}@/^1.5pc/[r]
\ar@{-}[r]
\ar@{-}@/_1.5pc/[r] 
& *{\bullet} \\} 
= \frac{1}{3}
\xymatrix{ *{\bullet} 
\ar@{-}@/^1.5pc/[r]^3
& *{\bullet} \\} 
+ \frac{1}{3}
\xymatrix{ *{\bullet} 
\ar@{-}[r]^3
& *{\bullet} \\} 
+ \frac{1}{3}
\xymatrix{ *{\bullet} 
\ar@{-}@/_1.5pc/[r]_3
& *{\bullet} \\} 
\]
Thus if the edges are labelled with kernels $K_{p_e}$ such that
$\int_M \, K_{p_e}(d_M(x,x_0))^3 \dx$ is absolutely convergent,
then the theta graph is absolutely convergent.
For example, this is true for the Lorentzian kernel,
for Cherrington's kernel with $k > 0$, and
for Pfeiffer's causal Lorentzian kernel, but is not true for 
Cherrington's kernel with $k = 0$ nor for Livine and Oriti's causal 
Lorentzian kernel.

Our proof that the $10j$ symbol is finite (Section~\ref{sec:10j})
was short because we made some clever choices of factor ordering 
and parentheses.
But we can also handle the $10j$ symbol following the pattern
of considering the spanning trees, analogous to what was done
above for the theta graph.
Each spanning tree of the $10j$ graph contains four of the ten edges.
By symmetry, each edge occurs in $4/10$ of the spanning trees.
So if we weight the edges in each spanning tree with weight $10/4$,
and take the convex linear combination of them having equal
coefficients $1/\text{(number of spanning trees)}$, 
they will add up to the $10j$ graph.
All of the above kernels are absolutely convergent for this weight.

Some graphs require a more subtle approach.
For example, consider the following graph:
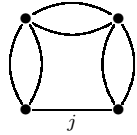
\[
\xymatrix@=2.5pc{*{\bullet} 
\ar@{-}@/^.5pc/[r]
\ar@{-}@/_.5pc/[r] 
\ar@{-}@/^.5pc/[d]
\ar@{-}@/_.5pc/[d] 
& *{\bullet} 
\ar@{-}@/^.5pc/[d]
\ar@{-}@/_.5pc/[d] 
\\ *{\bullet}
\ar@{-}[r]_j
& *{\bullet} \\} 
\]
This graph has 20 spanning trees, 12 of which contain the
edge labelled $j$.
So if these were combined using equal coefficients ($1/20$),
the weights assigned to the 12 trees containing $j$ would
have to be $20/12$ on average.  Since $20/12 \leq 2$,
at least one of these trees would be divergent with all
of the kernels considered in this paper.
However, consider the following 7 spanning trees:
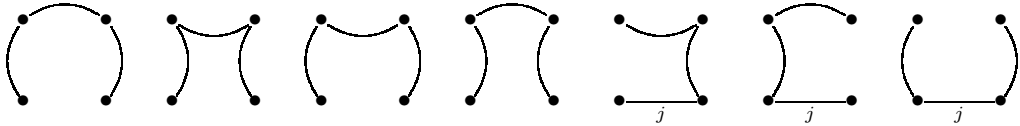
\[
\xymatrix@=2.2pc{*{\bullet} 
\ar@{-}@/^.5pc/[r]
\ar@{-}@/_.5pc/[d] 
& *{\bullet} 
\ar@{-}@/^.5pc/[d]
\\ *{\bullet}
& *{\bullet} \\}\qquad 
\xymatrix@=2.2pc{*{\bullet} 
\ar@{-}@/_.5pc/[r] 
\ar@{-}@/^.5pc/[d]
& *{\bullet} 
\ar@{-}@/_.5pc/[d] 
\\ *{\bullet}
& *{\bullet} \\}\qquad 
\xymatrix@=2.2pc{*{\bullet} 
\ar@{-}@/_.5pc/[r] 
\ar@{-}@/_.5pc/[d] 
& *{\bullet} 
\ar@{-}@/^.5pc/[d]
\\ *{\bullet}
& *{\bullet} \\}\qquad 
\xymatrix@=2.2pc{*{\bullet} 
\ar@{-}@/^.5pc/[r]
\ar@{-}@/^.5pc/[d]
& *{\bullet} 
\ar@{-}@/_.5pc/[d] 
\\ *{\bullet}
& *{\bullet} \\}\qquad 
\xymatrix@=2.2pc{*{\bullet} 
\ar@{-}@/_.5pc/[r] 
& *{\bullet} 
\ar@{-}@/_.5pc/[d] 
\\ *{\bullet}
\ar@{-}[r]_j
& *{\bullet} \\}\qquad 
\xymatrix@=2.2pc{*{\bullet} 
\ar@{-}@/^.5pc/[r]
\ar@{-}@/^.5pc/[d]
& *{\bullet} 
\\ *{\bullet}
\ar@{-}[r]_j
& *{\bullet} \\}\qquad 
\xymatrix@=2.2pc{*{\bullet} 
\ar@{-}@/_.5pc/[d] 
& *{\bullet} 
\ar@{-}@/^.5pc/[d]
\\ *{\bullet}
\ar@{-}[r]_j
& *{\bullet} \\} 
\]
Each edge occurs in exactly 3 of these 7 trees, so if
all edges are given weight $7/3$, then the sum of these
trees with all coefficients equal to $1/7$ gives the
original graph.
All of the kernels considered in this paper are absolutely
convergent for this weight.

There are graphs for which you have to be even more clever,
by choosing varying coefficients.
For example, one can use these methods to show that
the graph
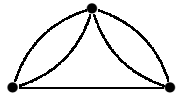
\[
\xymatrix{& *{\bullet} 
\ar@{-}@/_.5pc/[dl]
\ar@{-}@/^.5pc/[dl]
\ar@{-}@/_.5pc/[dr]
\ar@{-}@/^.5pc/[dr]
\\ *{\bullet} \ar@{-}[rr] & & *{\bullet}
}
\]
has convergent evaluations for all of the above
kernels, by using weights of $5/2$ and varying
coefficients (so the given graph isn't the barycenter
of the trees).

Finally, the method does not work for all graphs.
For example, for the $6j$ symbol, which was shown by
Baez and Barrett to be absolutely convergent for the
Lorentzian kernel~\cite{BaBa01}, it is not possible
to choose all of the weights strictly greater than 2
no matter what coefficients you choose.

\section{Conclusions and Further Work}
\label{sec:conc}

The main idea of this paper is to use convexity to obtain
bounds on integrands which can be expressed as products
of more easily understood factors.  The inequalities reduce the number
of factors, but in exchange increase the exponents on
the remaining factors.  
For some graphs and some kernels, we can ensure that
the exponents are in a range that makes the individual
kernels converge, and so one can deduce convergence of
the original integral.
This technique allows us to reprove old convergence results in a 
simpler way, and to easily prove new convergence results.

At present, the methods do not work for the Lorentzian 6j
symbol.  It would be interesting to figure out a way to
strengthen them to handle this case.
More generally, it is an open question to classify
which graphs give absolutely convergent Lorentzian evaluations.
One conjecture is that this are exactly the 3-connected 
graphs~\cite{BaBa01}.
Can the methods here be strengthened to prove this?
What can we say about the situation with other kernels
and other homogeneous domains?
For example,
it would also be useful to know whether the Euclidean 10j symbol
from~\cite{BCE}, which is based on the kernel $\sin(p r)/r$ and the
homogeneous domain $\R^3$, is absolutely convergent.  My belief is
that it is only conditionally convergent, but even this is not known.

\section*{Acknowledgements}

I would like to thank John Baez, Wade Cherrington, Igor Khavkine, 
Finnur L\'arusson and Josh Willis for very helpful conversations
about this work.

\end{document}